\documentclass[english]{article}
\usepackage[T1]{fontenc}
\usepackage[latin9]{inputenc}
\usepackage{geometry}
\usepackage{babel}
\usepackage{verbatim}
\usepackage{float}
\usepackage{amsmath}
\usepackage{amssymb}
\usepackage{graphicx}
\usepackage{setspace}
\usepackage[unicode=true,pdfusetitle,
 bookmarks=true,bookmarksnumbered=false,bookmarksopen=false,
 breaklinks=false,pdfborder={0 0 1},backref=false,colorlinks=false]
 {hyperref}

\makeatletter
\usepackage{comment}

\@ifundefined{showcaptionsetup}{}{%
 \PassOptionsToPackage{caption=false}{subfig}}
\usepackage{subfig}
\makeatother

\begin{document}

\global\long\def\bC{\mathbb{C}}%

\global\long\def\bE{\mathbb{E}}%

\global\long\def\bF{\mathbb{F}}%

\global\long\def\bK{\mathbb{K}}%

\global\long\def\bN{\mathbb{N}}%

\global\long\def\bP{\mathbb{P}}%

\global\long\def\bQ{\mathbb{Q}}%

\global\long\def\bR{\mathbb{R}}%

\global\long\def\bT{\mathbb{T}}%

\global\long\def\bZ{\mathbb{Z}}%

\global\long\def\cA{\mathcal{A}}%

\global\long\def\cB{\mathcal{B}}%

\global\long\def\cC{\mathcal{C}}%

\global\long\def\cD{\mathcal{D}}%

\global\long\def\cE{\mathcal{E}}%

\global\long\def\cF{\mathcal{F}}%

\global\long\def\cG{\mathcal{G}}%

\global\long\def\cH{\mathcal{H}}%

\global\long\def\cI{\mathcal{I}}%

\global\long\def\cJ{\mathcal{J}}%

\global\long\def\cK{\mathcal{K}}%

\global\long\def\cL{\mathcal{L}}%

\global\long\def\cLp#1{\mathcal{L}^{#1}}%

\global\long\def\cLpp#1{\mathcal{L}_{+}^{#1}}%

\global\long\def\cLsimp{\mathcal{L}_{simp}^{0}}%

\global\long\def\cM{\mathcal{M}}%

\global\long\def\cN{\mathcal{N}}%

\global\long\def\cO{\mathcal{O}}%

\global\long\def\cP{\mathcal{P}}%

\global\long\def\cR{\mathcal{R}}%

\global\long\def\cS{\mathcal{S}}%

\global\long\def\cT{\mathcal{T}}%

\global\long\def\cU{\mathcal{U}}%

\global\long\def\cV{\mathcal{V}}%

\global\long\def\cW{\mathcal{W}}%

\global\long\def\cY{\mathcal{Y}}%

\global\long\def\cZ{\mathcal{Z}}%

\global\long\def\fq{\mathscr{\mathfrak{q}}}%

\global\long\def\fH{\mathscr{\mathfrak{H}}}%

\global\long\def\sD{\mathscr{D}}%

\global\long\def\sH{\mathscr{H}}%

\global\long\def\sK{\mathscr{K}}%

\global\long\def\sF{\mathscr{F}}%

\global\long\def\norm#1{\left\Vert #1\right\Vert }%

\global\long\def\np#1#2{\left\Vert #1\right\Vert _{#2}}%

\global\long\def\nlp#1#2{\left\Vert #1\right\Vert _{L^{#2}}}%

\global\long\def\abs#1{\left|#1\right|}%

\global\long\def\inv#1{#1^{-1}}%

\global\long\def\adjoint#1{#1^{*}}%

\global\long\def\annihilator#1{#1^{\circ}}%

\global\long\def\annihilatee#1{#1^{\perp}}%

\global\long\def\unaryop#1{#1\left(\cdot\right)}%

\global\long\def\binaryop#1{#1\left(\cdot,\cdot\right)}%

\global\long\def\comp#1#2{#1\circ#2}%

\global\long\def\converge#1{\overset{#1}{\joinrel\longrightarrow}}%

\global\long\def\define{\triangleq}%

\global\long\def\enum#1#2{\left\{  #1_{1},\dots,#1_{#2}\right\}  }%

\global\long\def\enumvec#1#2{\left(#1_{1},\dots,#1_{#2}\right)}%

\global\long\def\enuminf#1{\left\{  #1_{1},#1_{2}\dots\right\}  }%

\global\long\def\equivalent{\Longleftrightarrow}%

\global\long\def\substitute#1{\overset{#1}{\joinrel===}}%

\global\long\def\tensor{\otimes}%

\global\long\def\liminf#1{\underset{#1}{\operatorname{lim\,inf}}}%

\global\long\def\limsup#1{\underset{#1}{\operatorname{lim\,sup}}}%

\global\long\def\essinf#1{\underset{#1}{\operatorname{ess\,inf}}}%

\global\long\def\esssup#1{\underset{#1}{\operatorname{ess\,sup}}}%

\global\long\def\sgn{\operatorname{sgn}}%

\global\long\def\spanset{\operatorname{span}}%

\global\long\def\Null{\operatorname{Null}}%

\global\long\def\Range{\operatorname{Range}}%

\global\long\def\io{\operatorname{i.o.}}%

\global\long\def\ae{\operatorname{a.e.}}%

\global\long\def\as{\operatorname{a.s.}}%

\global\long\def\d#1{\operatorname{d}#1}%

\global\long\def\D#1{\operatorname{D}#1}%

\global\long\def\Db#1{\operatorname{D}\left[#1\right]}%

\global\long\def\cov{\operatorname{cov}}%

\global\long\def\supp{\operatorname{supp}}%

\title{RLOP: RL Methods in Option Pricing from a Mathematical Perspective}
\author{Ziheng Chen\thanks{email: \protect\href{http://ziheng.chen@math.utexas.edu}{ziheng.chen@math.utexas.edu}}}
\maketitle
\begin{abstract}
In this work, we build two environments, namely the modified QLBS
and RLOP models, from a mathematics perspective which enables RL methods
in option pricing through replicating by portfolio. We implement the
environment specifications \footnote{the source code can be found at \href{https://github.com/owen8877/RLOP}{https://github.com/owen8877/RLOP}},
the learning algorithm, and agent parametrization by a neural network.
The learned optimal hedging strategy is compared against the BS prediction.
The effect of various factors is considered and studied based on how
they affect the optimal price and position.\footnote{We thank Zhou Fang for his initial motivation and helpful discussion
on the financial aspects.}
\end{abstract}

\section{Introduction and Background}

Option pricing is one of the central and open questions in the mathematical
finance field. It motivates the advancement of the stochastic process
theory and becomes one of the practical applications of sophisticated
theories. To begin with, options are a type of derivative that gives
the holder the right, but not the obligation, to buy or sell the underlying
asset at a specified price (a.k.a. the strike price) within a specific
period. The name `derivative' implies that those securities are derived
from the origin asset. Although investors can long and short options
for seeking risky profits, one primary feature of options is to obtain
a certain share of assets with a fixed amount of payment in the future.
This helps to control volatility and uncertainty. Thus, the market
maker who sells the option charges the buyer a certain amount of fee
that amounts to the expected risk.

We focus on the European call in this paper. One can only exercise
a European option at the terminal time and a call option gives the
holder the right to buy. The holder usually exercises the European
call if the asset price ends up above the strike price and takes no
action otherwise. If the market is liquid enough, the holder can sell
the asset immediately after obtaining it, making a marginal profit.
Therefore, we define the payoff of a European call as
\begin{equation}
\text{payoff}=\max\left\{ S-K,0\right\} \label{eq:intro-payoff}
\end{equation}
where $S$ is the asset price at the exercise time and $K$ stands
for the strike price. Although it is tempting to price the option
based on this payoff function with consideration of each path probability,
this is usually not the best way since the option is designed to measure
the risk of obtaining the asset. If there is a general trend in the
price process, one can hold a consistent position in the asset and
purchase an option in addition. Thus, the more popular approach is
to manage a portfolio that replicates the option, i.e. the holder
invests the same amount of money as the price of the option in a replication
portfolio and gets the same amount of payoff (Eqn. \ref{eq:intro-payoff})
at the terminal time. The portfolio borrows some extra funds from
an external risk-free source to hold a long position of the underlying
asset and adjusts its position according to the change in the asset
price. This approach is referred to as the replication by portfolio
method.

We conduct a brief review in this section, starting by introducing
the Black-Scholes model which solves the optimal hedging strategy
and the pricing problem at the same time, moving to a literature review
where multiple types of asset pricing are involved, and discussing
how to model transaction costs.

\subsection{Black-Scholes Model}

The Black-Scholes formula (or BS formula or short) is the first formula
that prices the European-style call option. There are two equivalent
ways of calculating the price, either by solving the Black-Scholes-Merton
equation or use the risk-neutral measure. The price of an European
call of time span $T$ and a strike price $K$ is solved by 
\begin{equation}
c_{\text{BS}}\left(t,x\right)=xN\left(d_{+}\left(T-t,x\right)\right)-Ke^{-r\left(T-t\right)}N\left(d_{-}\left(T-t,x\right)\right)\label{eq:bs-price}
\end{equation}
where $0\le t<T$ is the current time and $x$ is the current asset
price, with $r$ being the risk-free interest rate, $\sigma$ being
the volatility, $N\left(y\right):=\frac{1}{\sqrt{2\pi}}\int_{-\infty}^{y}\exp\left(-\frac{z^{2}}{2}\right)\d z$
being the cumulative standard normal distribution, and $d_{\pm}$
defined as
\[
d_{\pm}\left(\tau,x\right):=\frac{1}{\sigma\sqrt{\tau}}\left[\log\frac{x}{K}+\left(r\pm\frac{\sigma^{2}}{2}\right)\tau\right].
\]
The optimal hedging position can be derived as the partial derivative
of Eqn. \ref{eq:bs-price} as 
\begin{equation}
u_{\text{BS}}\left(t,x\right)=N\left(d_{+}\left(T-t,x\right)\right).\label{eq:bs-hedge}
\end{equation}
We leave the details to \cite{shreve2004stochastic,alma991057973644906011}
for more in-depth theory and computation. 

This set of theory surpassed most of the option pricing theories when
it was first published and is still helpful from since. With that
being said, the theory relies on a critical assumption that the agent
is able to continuously rehedge the positions of the replication portfolio.
This is either impractical or costly to most of the investing parties.
In the view of stochastic integration, we use a Riemann-style sum
with left endpoints to approximate the Ito integral. The accumulation
of the temporal discretization errors will affect the final payoff
profoundly.

\subsection{Literature Review}

There have been vast works and literatures on option pricing. A few
exercising strategies for American option are reviewed in \cite{li2009learning}.
The state space is composed of asset price trajectories $\left(S_{0},S_{1},\dots,S_{T}\right)$
and an absorbing state $\boldsymbol{e}$ which serves as the destination
after the option has been exercised. The action space simply contains
two actions: hold and exercise. The only non-zero reward is given
when the option has been exercised at the value specified by the option.
The Q-function is derived as in classical RL problems and it is the
quantity of interest. The first method, LSPI (LS policy iteration),
combines LSTD (LS with TD update) and policy iteration together to
achieve efficient learning. The second and third ones, FQI (fitted
Q-iteration algorithm) and LSMC (Least squares Monte Carlo), take
a DP approach where the exact exercising problem is solved at each
trading moment; the only difference lies in that FDI takes a forward
view in time while LSMC is backward, starting from the exercising
time. In the same spirit, \cite{fathan2021deep} investigates the
efficiency and effectiveness of different parametrizaiton structures.
Three algorithms, namely double deep Q-Learning (DDQN), categorical
distributional RL (C51), and implicit quantile networks (IQN) are
compared where DDQN learns the optimal Q-value function while the
other two algorithms try to learn the full distribution of the discounted
reward. They are tested on empirical data as well as simulated geometric
Brownian motion trajectories.

Different RL architectures can be deployed in this field as well.
The Actor-Critic structure is used in \cite{marzban2021deep} for
Equal Risk Pricing (ERP) in a risk averse setting under the framework
studied in \cite{tamar2015policy}. The concept of $\tau$-expectile
\[
\overline{\rho}\left(X\right)=\arg\min_{q}\tau\bE\left[\left(q-X\right)_{+}^{2}\right]+\left(1-\tau\right)\bE\left[\left(q-X\right)_{-}^{2}\right]
\]
is used to elicit a coherent risk measure. The value function is defined
as the portfolio value under the recursive coherent risk measure realized
by expectiles, i.e.
\[
V_{t}\left(S_{t},Y_{t}\right)=\inf_{\xi_{t}}\overline{\rho}\left(-\xi_{t}^{T}\Delta S_{t+1}+V_{t+1}\left(S_{t+1},Y_{t+1}\right)|S_{t},Y_{t}\right)
\]
with terminal condition $V_{T}\left(S_{T},Y_{T}\right)=F\left(S_{T},Y_{T}\right)$
specified by the option contract. With that being established, we
can apply the policy gradient method where the critic network updates
the value estimate which the actor network can refer to and build
its policy upon. The network uses a classical fully multilayer structure
with alternating activation functions.

A hybrid attempt is carried out in \cite{grassl2010reinforcement}
where the pricing strategy is based on a combination of optimal stopping
and terminal payoff. The idea is that the agent can either hold the
derivative until the terminal time, executing the contract to get
the payoff written, or sell the derivative earlier according to the
price at that particular moment. The reward function is thus simplified
defined as the selling/execution price if such scenario happens. The
value function is parameterized by kernel function approximation and
the algorithm is tested using simulated geometric Brownian paths of
an European call option.

We point out that it is also possible to directly solve the HJB equation
if the associated RL problem is formulated as a control porblem. We
refer to \cite{halperin2021distributional} for more details on distributional
offline continuous-time RL learning algorithms.

\subsection{Trading Cost \label{subsec:Trading-Cost}}

One other important factor in option pricing is the trading cost (or
transaction cost, used interchangeably). The trading cost occurs when
the hedging position of the replication portfolio changes, making
a portion of the invested capital unavailable. Thus, the presence
of trading cost increases the price of the option in general.

A way of understanding and modeling the trading cost is to use the
concept of bid-ask spread. Consider the limit order book of a particular
asset at a given time. Empirical observation shows that the selling
(a.k.a. ask) orders are always above the buying orders (a.k.a. bid)
orders since one wishes to buy low and sell high. Moreover, the orders
are not uniformly distributed: there are fewer orders close to the
mid price (i.e. the arithmetic average of the highest bid and lowest
ask) and more orders away from the mid price. The gap between the
lowest ask and highest bid is called the bid-ask spread. A larger
gap usually implies a higher friction in the market.

In general, the dynamics of the limit order book distribution is a
very challenging problem. However, under the assumption that the replicating
portfolio hedges a small position, we can safely conclude that the
trading cost, defined as the additional cost paid for buying (or selling,
correspondingly) a unit of asset with reference to its mid price,
can be computed as follows
\[
\text{Trading cost (TC)}=S_{\text{ask}}-S_{\text{mid}}=\frac{1}{2}\left(S_{\text{ask}}-S_{\text{bid}}\right)=\frac{1}{2}\text{spread}.
\]
To model the spread, we assume that it scales with the mid price by
a factor of $\epsilon$, characterizing the friction. Thus, the trading
cost to buy/sell $\Delta u$ shares of the asset requires a trading
cost of
\[
\text{TC}\left(\Delta u,S_{\text{mid}}\right)=\frac{\epsilon}{2}S_{\text{mid}}\Delta u.
\]

\section{QLBS: Q-learning Black-Scholes Model}

\subsection{A Brief Review}

We quickly introduce the QLBS model proposed in \cite{halperin2020qlbs}.
Consider a sequence of asset prices $\left\{ S_{t}\right\} _{t=0,1,\dots}$,
adapted under the filtration $\left\{ \cF_{t}\right\} $, upon which
we wish to build an option with payoff function $h$ at the maturity
time $T$. The option is realized as a hedge portfolio which consists
of some shares $u_{t}$ of the underlying asset and the risk-free
deposit $B_{t}$. Let 
\[
\Pi_{t}:=u_{t}S_{t}+B_{t}
\]
denote the value of the portfolio at time $t$. To fulfill the option
contract, the holding position is cleared at the terminal time $T$
and is fully converted into cash position, i.e.
\[
\Pi_{T}=B_{T}=h\left(S_{T}\right).
\]
The deposit position at a particular time is solved via the self-financing
condition which requires that the instantaneous value of the portfolio
is kept same before and after the re-hedging operation:
\begin{equation}
u_{t}S_{t+1}+e^{r\Delta t}B_{t}=u_{t+1}S_{t+1}+B_{t+1}\label{eq:self-financing}
\end{equation}
where $r$ stands for the risk-free interest rate. Eqn. \ref{eq:self-financing}
can be used to solve how much money is needed to cover future trading
activities.

A state is defined as a pair of an integer and a real number $\left(t,X_{t}\right)$,
where 
\[
X_{t}=-\left(\mu-\frac{\sigma^{2}}{2}\right)t+\log S_{t}
\]
is the compensated logarithm price. The action space is the real number,
indicating how much shares are hedged at a particular time. A policy
is a mapping from the state space to the action space, i.e.
\[
\pi:\left(t,X_{t}\right)\mapsto a_{t}.
\]
Notice that we use $a_{t}$ for the log-processed input $X_{t}$ while
$u_{t}$ for the asset price $S_{t}$ in normal scale. The policy
may depend on other macro factors, e.g. interest rate $r$, volatility
$\sigma$, total maturity time $T$, the strike price $K$ if the
option is of call/put type.

The reward function is derived from the Bellman's optimality equation
where we define the value function in the first place. The idea is
to minimize the money needed to initiate the hedge portfolio as well
as to minimize the volatility throughout the trading periods. Given
a hedging strategy $\pi$, the value function is defined as
\begin{equation}
V_{t}^{\pi}\left(X_{t}\right)=\bE^{\pi}\left[-\Pi_{t}\left(X_{t}\right)-\lambda\sum_{\tau=t}^{T}e^{-r\left(\tau-t\right)}\text{Var}\left[\Pi_{\tau}\left(X_{\tau}\right)|\cF_{\tau}\right]|\cF_{t}\right]\label{eq:qlbs-value-function}
\end{equation}
where $\lambda$ is the risk aversion factor. The reward function
can be derived by matching the corresponding terms in the Bellman's
equation:
\[
R_{t}\left(X_{t},a_{t},X_{t+1}\right):=\gamma a_{t}\Delta S_{t}-\lambda\text{Var}\left[\Pi_{t}|\cF_{t}\right]
\]
where $\gamma:=e^{-r\Delta t}$ is the discounting factor. The connection
between the value function and option pricing is that the option price
is given by the minus optimal Q-function.

\subsection{Make QLBS Interactive}

Elegant as the vanilla QLBS approach, it is not a true RL problem
since the optimal policy $\pi^{*}$ is solved analytically without
any reinforcement learning techniques. In fact, the author derives
the Bellman's equation for the optimal Q-function from Eqn. \ref{eq:qlbs-value-function}
\begin{equation}
Q_{t}^{*}\left(X_{t},a_{t}\right)=\gamma\bE_{t}\left[Q_{t+1}\left(X_{t+1},a_{t+1}^{*}\right)+a_{t}\Delta S_{t}\right]-\lambda\gamma^{2}\bE_{t}\left[\widehat{\Pi}_{t+1}^{2}-2a_{t}\widehat{\Pi}_{t+1}\Delta\widehat{S}_{t}+a_{t}^{2}\left(\Delta\widehat{S}_{t}\right)^{2}\right]\label{eq:qlbs-optimal-Q-function}
\end{equation}
which admits the optimal policy in closed form since Eqn. \ref{eq:qlbs-optimal-Q-function}
is a quadratic function in $a_{t}$. Such direct approach is feasible
if provided with abundant data on the correlation structure of the
portofolio value $\widehat{\Pi}$ and the stock price change $\Delta\widehat{S}$,
but it fails to generalize beyond this simple setting. Besides, the
portofolio value process $\Pi_{t}$ is in general non-adapted due
to how the self-financing condition (Eqn. \ref{eq:self-financing})
works. 

\begin{figure}[H]
\begin{centering}
\includegraphics[width=0.6\linewidth]{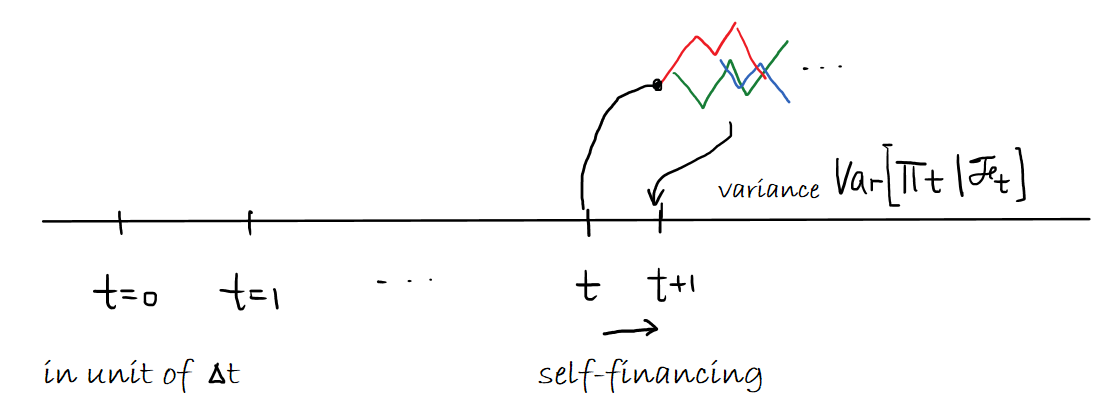}
\par\end{centering}
\caption{A brief illustration on the modified QLBS environment.}

\label{fig:qlbs-illustration}
\end{figure}

To deal with the aforementioned issues, we propose a modified QLBS
model which is
\begin{enumerate}
\item fully adapted with respect to the given filtration $\left\{ \cF_{t}\right\} $,
\item compatible with the transaction cost proposed in Sec. \ref{subsec:Trading-Cost},
and
\item works well with value-based or policy-based learning algorithms.
\end{enumerate}
We start by modifying the value function. A first problem lies in
the fact that the reward are not homogeneous over time. Eqn. \ref{eq:qlbs-value-function}
assigns the (negative) cashflow with a risk part as the reward, but
the terminal step gets the option payoff which is in general much
larger than the previous steps. Technically speaking, an agent could
notice this heterogenity since the time $t$ aligns with the terminal
time $T$, but it is rather difficult in practice to figure out this
situation. Thus, we propose the modified value function 
\begin{equation}
V_{t}^{\pi}\left(X_{t}\right)=\bE_{t}^{\pi}\left[-\left(1-\frac{t}{T}\right)\Pi_{t}\left(X_{t}\right)-\lambda\sum_{\tau=t}^{T}\gamma^{\tau-t}\sqrt{\text{Var}\left[\Pi_{\tau}\left(X_{\tau}\right)\right]}\right].\label{eq:qlbs-modified-V}
\end{equation}
Our improvement is two folds:
\begin{enumerate}
\item We weight the portfolio term by a diminishing factor $\left(1-\frac{t}{T}\right)$.
This factor does not impact the starting estimate at $t=0$ and it
fully vanishes at $t=T$. We point out that introducing this factor
will break the temporal symmetry so that the intermediate estimate
does not correspond to option pricing from the intermediate time steps.
\item We take a square root of the variance terms so that they turn into
standard deviation. This helps to keep the value estimate dimensionless
and robust.
\end{enumerate}
The reward function changes accordingly to
\begin{equation}
R_{t+1}\left(X_{t},a_{t}\right)=\bE_{t}^{\pi}\left[-\left(1-\frac{t}{T}\right)\Pi_{t}\left(X_{t}\right)+\left(1-\frac{t+1}{T}\right)\Pi_{t}\left(X_{t}\right)-\lambda\sqrt{\text{Var}\left[\Pi_{t}\left(X_{t}\right)\right]}\right]\label{eq:qlbs-modified-reward}
\end{equation}
with action at time $t$ to be $a_{t}$ and remaining actions following
the current policy $\pi$. Considering the effect of transaction costs,
the portfolio value process $\Pi_{t}$ is calculated backwards from
the modified self-financing condition
\begin{equation}
e^{r\Delta t}\left(\Pi_{t}-u_{t}S_{t}\right)+u_{t}S_{t+1}=\Pi_{t+1}+\text{TC}\left(u_{t+1}-u_{t},S_{t+1}\right)\label{eq:qlbs-modified-self-financing}
\end{equation}
with terminal condition $\Pi_{T}=h\left(S_{T}\right)$ unchanged.
We point out that we wrap the cashflow part with an conditional expectation
at time $t$ so that we don't run into adaptedness issues. We briefly
illustrate the modified QLBS model in Fig. \ref{fig:qlbs-illustration}.

\subsection{Specification}

In this section, we describe how to set-up the QLBS environment and
the necessary numerical procedures.

\subsubsection{Environment \label{subsec:Environment}}

The environment is responsible for keeping track of the asset price
and portfolio value based on the actions provided by the agent. With
a given set of parameters $r,\mu,\sigma,T$, the asset prices are
a set of geometrical brownian motion paths $\left\{ S_{t}\right\} $
which solves
\[
\d{S_{t}}=\mu S_{t}\d t+\sigma S_{t}\d{W_{t}}
\]
where $W_{t}$ refers to the standard brownian motion according to
the filtration $\left\{ \cF_{t}\right\} $. At each time step $t$,
the normalized price $X_{t}=-\left(\mu-\frac{\sigma^{2}}{2}\right)t+\log S_{t}$
and $t$ are provided to the agent, waiting for the response of the
hedge position $a_{t}$. Then, the reward $R_{t+1}$, as a conditional
expectation specified in Eqn. \ref{eq:qlbs-modified-reward}, is computed
empirically by averaging samples from a fixed number of additional
trajectories under the current policy. At the beginning of each episode,
the parameters are adjusted in a random fashion to help the agent
explore different settings and help avoid overfitting; the adjustment
obeys a Poisson process with intensity $\varUpsilon$.

\subsubsection{Agent Parametrization \label{subsec:Agent-Parametrization}}

\begin{figure}[H]
\begin{centering}
\includegraphics[width=0.8\linewidth]{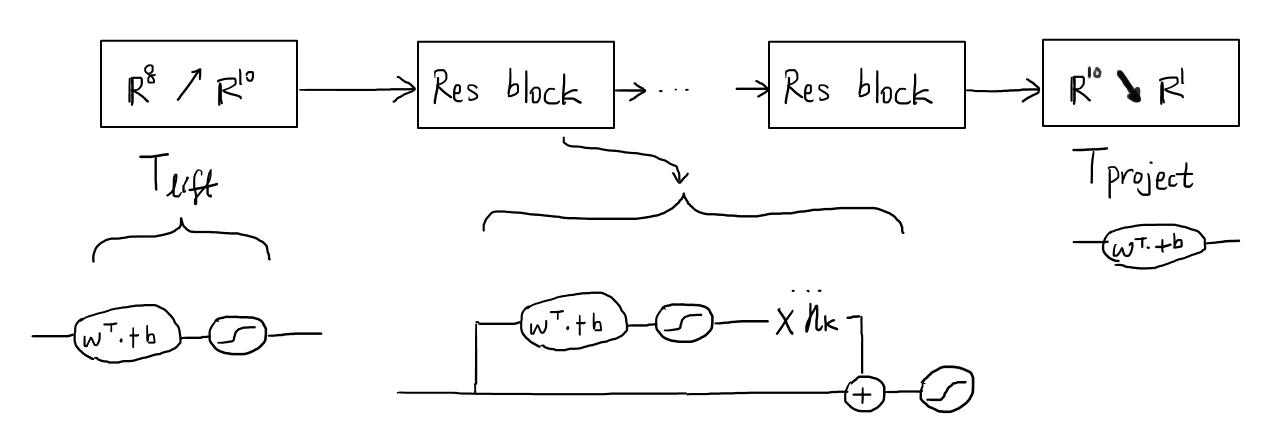}
\par\end{centering}
\caption{An illustration on a Resnet-motivated network structure.}

\label{fig:qlbs-resnet}
\end{figure}

The agent is fully responsible for determining the hedge position
$a_{t}$ under a given normalized price at each time step. The policy
$\pi$, whether stochastic or deterministic, depends on these input
variables as well as the environment parameters
\[
a_{t}\sim\pi\left(X_{t},t;r,\mu,\sigma,T,K,\lambda\right).
\]
In pratice, we prefer a stochastic policy since it encourages exploration
which is helpful to excape local minima. To further simplify the sampling
procedure, we restrict our policy spaces to Gaussian distribution
where the agent determines the mean and standard deviation, i.e.
\[
\pi=\cN\left(\mu_{\pi},\sigma_{\pi}\right)
\]
(where the subscripts $\pi$ are used to distinguish these parameters
from the environmental ones). The statistics $\mu_{\pi}$ and $\sigma_{\pi}$
are parametrized by two separate neural networks with the Resnet \cite{he2016deep}
skip-connection structure. The Resnet structure is composed of three
parts:
\begin{enumerate}
\item Pre-processing $T_{\text{lift}}\left(x\right):=\Xi\left(w_{\text{lift}}^{T}x+b_{\text{lift}}\right)$,
that lifts the (8-dimensional) input to the latent dimension by an
affine transform and an activation funciton $\Xi$;
\item Chain of transforms $\left\{ T^{\left(k\right)}\right\} $ in a fixed-point
iteration style, with each transform combining the identity functions
and a series of alternating affine transforms $\left\{ Z_{l}^{\left(k\right)}\right\} $
and activation $\Xi$, i.e.
\[
T^{\left(k\right)}:=\Xi\circ\left[\text{id}+Z_{n_{k}}^{\left(k\right)}\circ\Xi\circ Z_{n_{k}-1}^{\left(k\right)}\circ\cdots\circ\Xi\circ Z_{1}^{\left(k\right)}\right]
\]
\item Post-processing $T_{\text{project}}\left(x\right):=w_{\text{project}}^{T}x+b_{\text{project}}$,
that projects the latent representations onto the target space.
\end{enumerate}
Thus, the Resnet realization can be formally written as $T_{\text{project}}\circ T^{\left(k\right)}\circ T^{\left(k-1\right)}\circ\cdots\circ T^{\left(1\right)}\circ T_{\text{lift}}$.
We include an illustration in Fig. \ref{fig:qlbs-resnet}.

\subsubsection{Learning Algorithm}

Since the action space $a_{t}\in\bR$ is a continuous space, it is
nature to adopt a policy-based method. Here we opt-in the classical
REINFORCE algorithm with baseline where the policy components relies
on the value estimator to learn quickly and reliably while the value
estimator learns from empirical averages of Monte Carlo samples. The
value estimator, parameterized by a neural network, uses the same
Resnet structure as mentioned in the previous section. We refer to
\cite{sutton2018reinforcement} for more details.

In our implementation, the policy network and the value network have
the same latent dimension 10 and they are composed of two Resnet blocks
with two hidden affine transforms. We use the Adam optimizers to update
the networks, one for each, and the learning rate is set to $10^{-4}$
by default.

\subsection{Experiments and Results}

We include a variety of experiments, starting from a demonstration
that shows the agent learns over time, moving to a comparison with
the Black-Scholes baseline model, and finally exploring other directions
that have a stronger connection to finance.

\subsubsection{Demonstration on learning\label{subsec:Demonstration-on-learning}}

\begin{figure}[H]
\subfloat[No adjustment on environment parameters. The shade indicates 95\%
confidence interval. Learning rate set to $10^{-4}$.]{\includegraphics[width=0.45\linewidth]{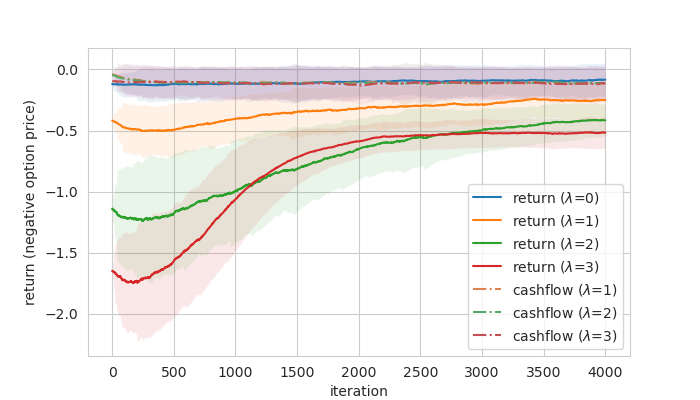}

\label{fig:qlbs-exp1-learning}}\qquad{}\subfloat[Adjustment parameter (on the initial price) $\varUpsilon=0.005$,
indicated by purple lines. The shade indicates 95\% confidence interval.
Learning rate set to $10^{-4}$. ]{\includegraphics[width=0.45\linewidth]{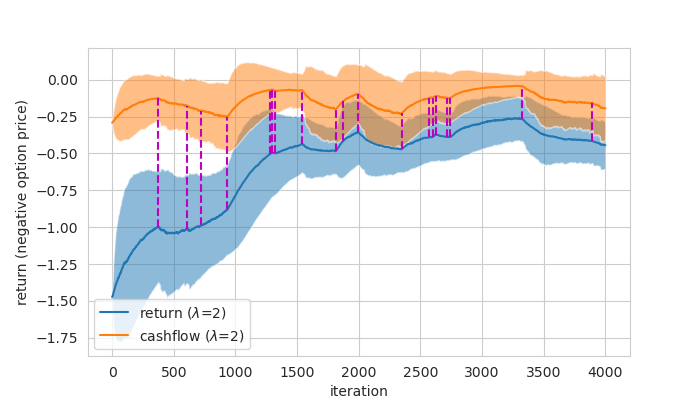}

\label{fig:qlbs-exp1-mutate}}

\caption{Exponentially moving average of episodic return.}

\label{fig:qlbs-exp1}
\end{figure}

We start by showing that the policy gradient algorithm has been correctly
implemented and the agent learns well over time. The environment parameter
is set to $r=0.01,\sigma=0.1,\mu=0,T=5,K=1,S_{0}=1,\Delta t=1,\lambda\in\left\{ 0,1,2,3\right\} ,\epsilon=0$
in Sec. \ref{subsec:Demonstration-on-learning}, \ref{subsec:Influence-of-risk},
and \ref{subsec:Effect-of-transaction} unless otherwise specified.
As shown in Fig. \ref{fig:qlbs-exp1-learning}, the episodic return
flatterns after training for around 3000 steps. It is worth noticing
that the cashflow return part 
\[
\sum_{t=0}^{T-1}\bE^{\pi}\left[-\left(1-\frac{t}{T}\right)\Pi_{t}\left(X_{t}\right)+\left(1-\frac{t+1}{T}\right)\Pi_{t}\left(X_{t}\right)\right]=\bE^{\pi}\left[\Pi_{0}\left(X_{0}\right)\right]
\]
is slightly smaller for a large risk parameter $\lambda$, implying
that the agent learns to make a trade-off between the cash-flow and
the risk component. 

To prepare for learning in a much broader setting, we allow the initial
price to be adjusted at a given intensity $\varUpsilon$. We mark
out the episodes that the adjustments take place (Fig. \ref{fig:qlbs-exp1-mutate})
and it seems that the agent can swiftly adapt to the change.

\subsubsection{Influence of risk parameter $\lambda$\label{subsec:Influence-of-risk}}

\begin{figure}[H]
\begin{centering}
\includegraphics[width=0.75\linewidth]{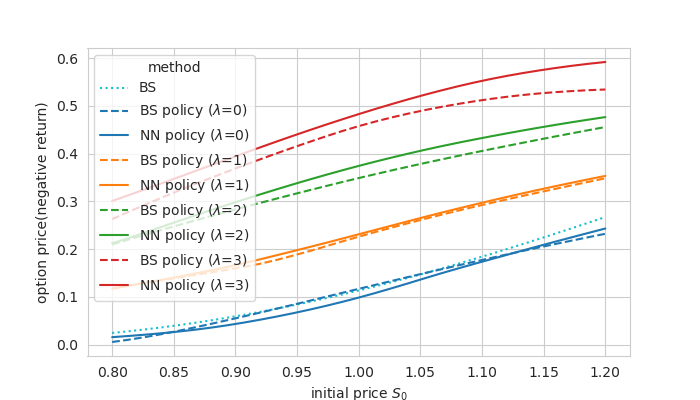}
\par\end{centering}
\caption{Negative episodic returns as option prices.}

\label{fig:qlbs-exp2}
\end{figure}

We wish to compare the optimal price under different choices of the
risk aversion parameter $\lambda$. A priori estimate is that a larger
$\lambda$ leads to a higher price learnt, since it penalizes improper
hedges harder. To complete the comparision, we also introduce the
vanilla BS hedging strategy $\pi_{\text{BS}}$ and learns the episodic
return using the same neural network parametrization. The results
are shown in Fig. \ref{fig:qlbs-exp2}. We remind the readers that
the option prices are defined as the negative value function, so a
lower option price implies a larger return learnt. As we expect, a
large $\lambda$ does lead to a higher price, both under BS policy
and under the learnt policy. In general, the policy learnt by the
neural network parametrization is a bit sub-optimal compared to the
BS policy, but it performs better when $\lambda=0.$

\subsubsection{Effect of transaction cost\label{subsec:Effect-of-transaction}}

\begin{figure}[H]
\begin{centering}
\includegraphics[width=1\linewidth]{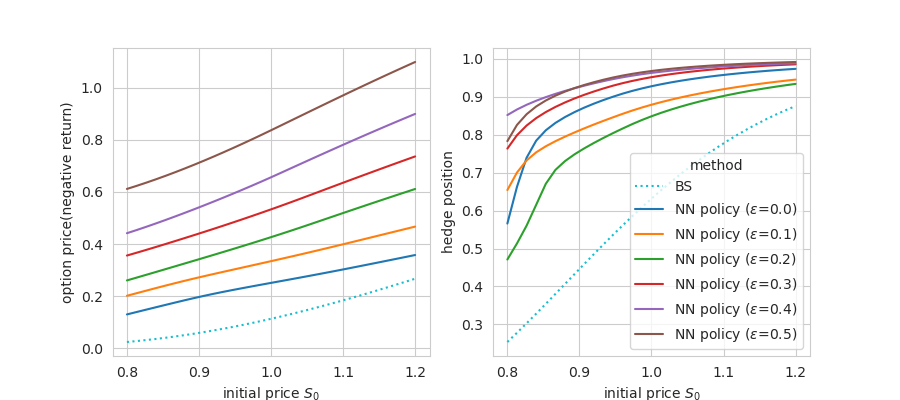}
\par\end{centering}
\caption{Option priced and hedged under different transaction cost parameters
$\epsilon$. $\lambda$ is set to 0.5 to introduce some risk concerns.}

\label{fig:qlbs-exp3}
\end{figure}

Next, we'd like to examine the effect of transaction costs. As indicated
by the self-financing condition (Eqn. \ref{eq:qlbs-modified-self-financing}),
a higher trading cost can significantly bias the portfolio value process
and consequently increase the optimal price learnt. As shown in Fig.
\ref{fig:qlbs-exp3}, we compare the price and hedge position learnt
under different friction parameters $\epsilon$. In general, a larger
$\epsilon$ does lead to a higher price learned (shown in the left
half) and the hedge position is usually always higher (in the right
half). 

\subsubsection{Generalization power}

\begin{figure}[H]
\begin{centering}
\includegraphics[width=1\linewidth]{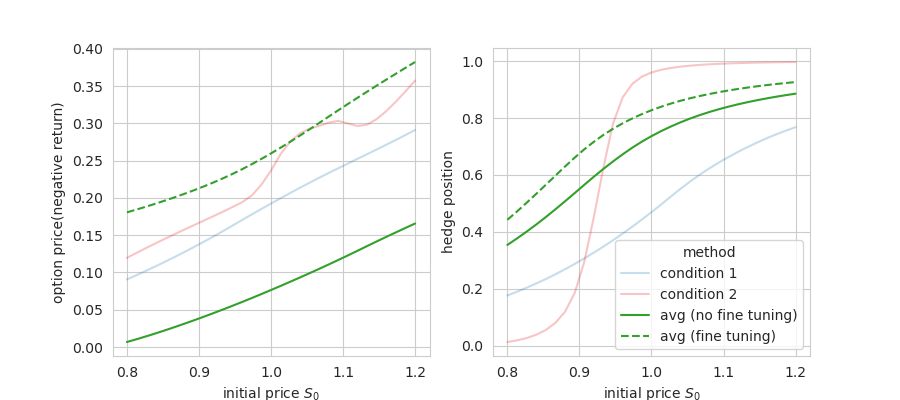}
\par\end{centering}
\caption{Option priced and hedged under different conditions.}

\label{fig:qlbs-exp4}
\end{figure}

As a concluding setting, we wish to parametrize the policy and baseline
networks on not only the states but also the environment parameters
and examine the generalization power. We pick two set of parameters:
$r=0.01,\mu=0,\sigma=0.1,\lambda=0.5,\epsilon=0$ under the first
condition while $r=0.02,\mu=0.1,\sigma=0.2,\lambda=1.5,\epsilon=0.1$
under the second condition. We train a policy and baseline under these
two conditions mixed, i.e. they have the same probability of being
presented, with the switching process being a Poisson process. After
training for a while, we refine the agent to learn a third condition
where the parameters are set to the arithmic average of the given
two conditions. As shown in Fig. \ref{fig:qlbs-exp4}, we examine
the hedging strategy of the agent before and after fine tuning to
the third average condition. The result, that the fine tuning does
a great improvement, is not surprising since we don't expect a lot
of generalization power. However, one could expect such generalization
given enough amount of training time.

\section{RLOP: Replication Learning of Option Pricing}

In this section, we propose a novel algorithm that prices a call/put
option via portfolio replication, but this method uses a forward view
which is fundamentally different from the QLBS approach. The idea
is simple: the agent manages a portfolio which yields a reward at
the terminal time based on how accurate the portfolio value is compared
to the option payoff. This naive idea has the problem that the reward
is zero for quite a long time until the maturity, which is usually
not good for shaping the agent's behavior. To deal with this downside,
we propose to group a few options as an ensemble so that the agent
gets a stream of feedbacks during each episode. To be specific, given
a (simulated or historical) path of the asset price $\left\{ S_{t}\right\} $,
maturity time $T$, and the payoff function $h$, the agent needs
to manages (at most) $T$ portfolios $\left\{ \Pi_{t}^{\left(i\right)}\right\} _{i=1}^{T}$
at the same time, where the $k$-th portfolio replicates the option
that terminals at time step $k$. In other words, for every $i\in\left[t+1,T\right]$,
the agent proposes the hedge position $u_{t}^{\left(i\right)}$ based
on the current time step $t$, the asset price $S_{t}$, and the balance
of the $i$-th portfolio $\Pi_{t}^{\left(i\right)}$. We hold the
belief that the agent is able to learn the hedging strategy step by
step from small $t$ to the terminal time $T$.

\subsection{MDP Formulation}

\begin{figure}[H]
\begin{centering}
\includegraphics[width=0.6\linewidth]{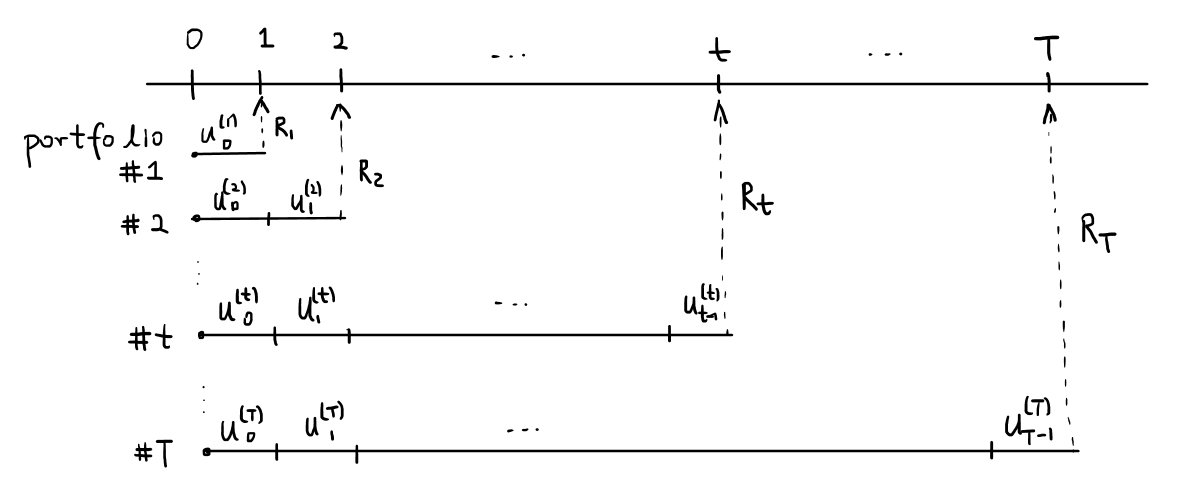}
\par\end{centering}
\caption{A brief illustration on the RLOP environment.}

\label{fig:rlop-illustration}
\end{figure}

We now rigorously define the aforementioned problem as a MDP. Given
a maturity time $i$, the state space consists of tuples containing
the time step $t$, the asset price $S_{t}$, and the current portfolio
value $\Pi_{t}^{\left(i\right)}$. The action space is $\bR$ that
contains all possible hedging positions. The transitional probability
(density) function is defined as
\[
p\left(\left(t,S_{t},\Pi_{t}^{\left(i\right)}\right),u_{t}^{\left(i\right)}\to\left(t',S_{t'},\Pi_{t'}^{\left(i\right)}\right),R_{t+1}\right)=\begin{cases}
\delta_{t+1,t'}\delta_{\widetilde{\Pi}_{t+1}^{\left(i\right)},\Pi_{t'}^{\left(i\right)}}\rho\left(S_{t},S_{t'}\right) & t<i\\
0 & t=i
\end{cases}
\]
where 
\begin{itemize}
\item the only admissible state is the terminal state when $t>i$, marking
the end of this episode;
\item $\widetilde{\Pi}_{t+1}^{\left(i\right)}$ refers to the portfolio
value determined by the self-financing condition
\[
\widetilde{\Pi}_{t+1}^{\left(i\right)}=e^{r\Delta t}\left(\Pi_{t}-u_{t}^{\left(i\right)}S_{t}\right)+u_{t}^{\left(i\right)}S_{t+1}-\text{TC}\left(u_{t+1}^{\left(i\right)}-u_{t}^{\left(i\right)},S_{t+1}\right)
\]
(which is the same as Eqn. \ref{eq:qlbs-modified-self-financing});
\item $\rho$ characterizes the dynamics of the underlying asset, e.g. the
discrete version geometric brownian motion;
\item $R_{t+1}=0$ for $t+1<i$ and $R_{i}=H\left(h\left(S_{i}\right),\Pi_{i}^{\left(i\right)}\right)$
where $H$ is a given penalty function that measures how accurate
the portfolio value $\Pi_{i}^{\left(i\right)}$ mimics the option
payoff $h\left(S_{i}\right)$.
\end{itemize}
As we mentioned in the previous paragraph, we stack a few pricing
tasks together so that the agent gets a reward from the portfolio
that recently terminates. For a $T$-term option, we combine $T$
tasks together where the agent is required to learn to price not only
the $T$-term one but all $t$-term ones for $t<T$. We briefly illustrate
the stucture of RLOP in Fig. \ref{fig:rlop-illustration}.

\subsection{Specification}

In this section, we describe how to set-up the RLOP environment and
the necessary numerical procedures.

\subsubsection{Environment}

As in the QLBS model, the environment samples a family of asset price
trajectories and maintains the portfolio value process. The asset
value is assumed to obey the geometrical brownian motion, as in Sec.
\ref{subsec:Environment}. We also allows the environment parameters
to change at the beginning of each episode according to a Poisson
process with intensity $\varUpsilon$.

\subsubsection{Agent Parametrization and Learning Algorithm}

The agent, i.e. the hedging strategy is parametrized by a Resnet-motivated
structure, same in Sec. \ref{subsec:Agent-Parametrization}. The REINFORCE
algorithm is adopted to train the agent. One caveat is that there
is no need to use a baseline as in the QLBS model, since in this case
the reward function is for penalty rather than learning the optimal
price.

\subsection{Experiments and Results}

We include a variety of experiments, including a demonstration that
shows the agent learns over time, a comparison with the Black-Scholes
baseline model, and a another one on the effect of trading costs.

\subsubsection{Demonstration on learning\label{subsec:Demonstration-on-learning-rlop}}

\begin{figure}[H]
\subfloat[Episodic return improves during training. Top figure: no adjustment
on the initial price $S_{0}$; bottom figure: the initial price is
adjusted via a Poisson process.]{\includegraphics[width=0.47\linewidth]{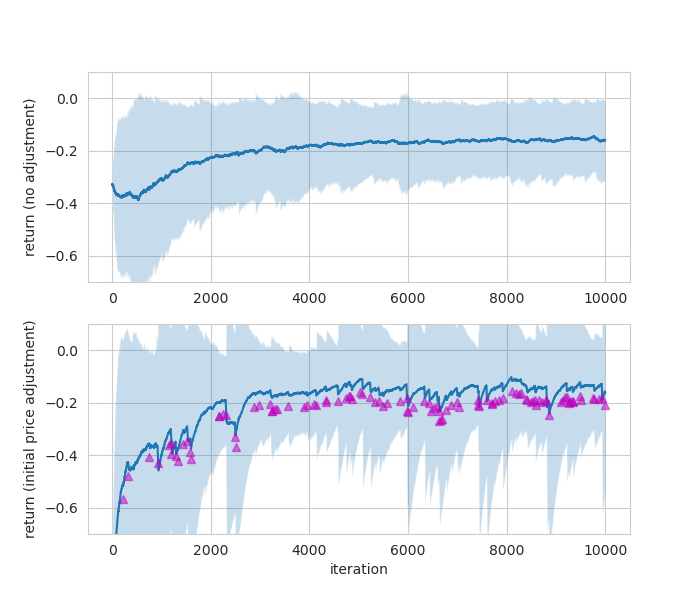}

\label{fig:rlop-exp1-learns}}\qquad{}\subfloat[Optimal hedge position learnt compared with BS predictions (Eqn. \ref{eq:bs-hedge}).]{\includegraphics[width=0.47\linewidth]{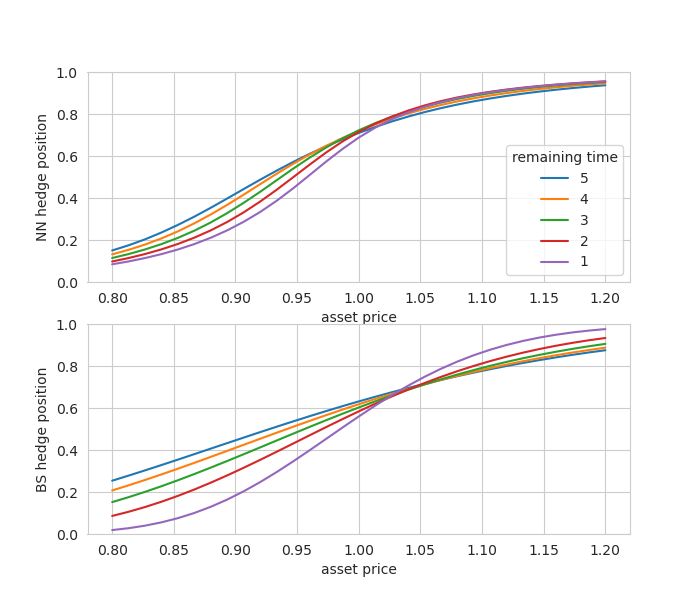}

\label{fig:rlop-exp1-hedge}}

\caption{Demonstration the RLOP model learning curve and a comparison to the
BS predictions.}

\label{fig:rlop-exp1}
\end{figure}

We use a parameter setting similar to Sec. \ref{subsec:Demonstration-on-learning}
where $r=0.01,\sigma=0.1,\mu=0,T=5,K=1,S_{0}=1,\Delta t=1$. The agent
is trained under no transaction cost with a fixed initial asset value
and the training curve is flat after around 8000 steps, shown in Fig.
\ref{fig:rlop-exp1-learns}. Similar to the QLBS setting, we introduce
adjustment to the parameters (i.e. the initial asset value in the
experiments to follow) by a Poisson process. We mark where the adjustment
occurs in the bottom half of the figure by purple triangles.

We also compare the optimal hedging strategy learnt with the BS predictions.
In general, the hedging curve should have a greater slope for a smaller
remaining time which implies that the price of the option is more
sensitive to the underlying asset price. As expected, both the learnt
position and the BS position show a greater slope when approaching
the terminal time. The one learnt by our RL agent is slightly insensitve
to the time variable.

\subsubsection{Effect of transaction cost\label{subsec:Effect-of-transaction-rlop}}

\begin{figure}[H]
\includegraphics[width=1\linewidth]{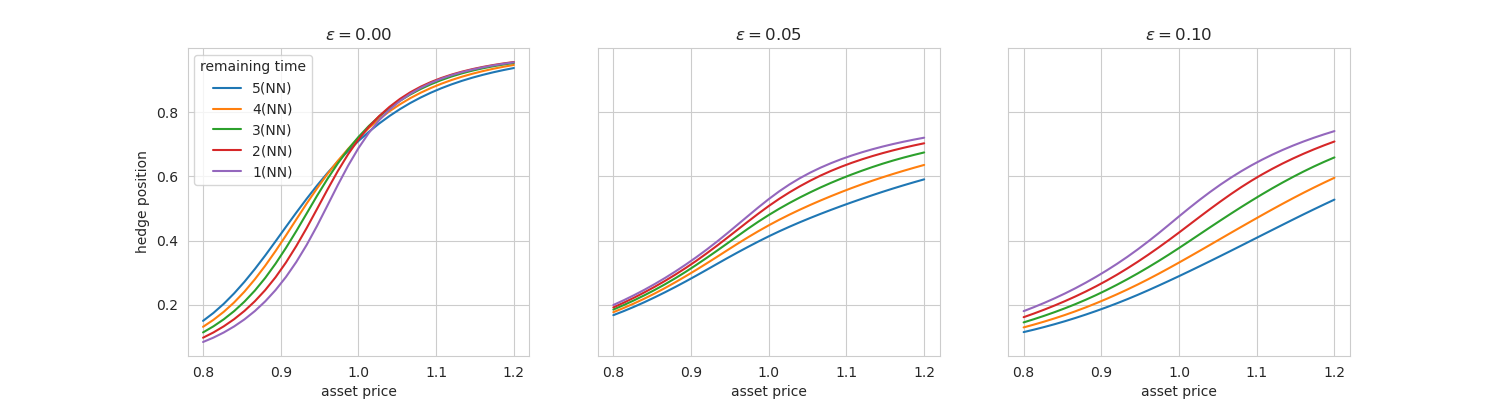}

\caption{Optimal hedging strategy learnt under different transaction costs.}

\label{fig:rlop-exp2}
\end{figure}

We proceed to measure how the transaction cost can effect the optimal
hedging strategy. We use the same setting as laid out in Sec. \ref{subsec:Effect-of-transaction}
where $\epsilon$ is the friction parameter. We plot the optimal hedging
position learnt under different $\epsilon$ in Fig. \ref{fig:rlop-exp2}.
The general trend is that a large $\epsilon$ discourages hedging
amounts, since it leads to a higher trading cost on average. A higher
friction also leads to a larger distinction between positions at different
times, which implies that in an illuquid market, we need to value
the time variable more than in a liquid one.

\section{Conclusion and Discussion}

In this project, we construct the modified QLBS and RLOP as two mathematically
sound environments that enable RL methods in the field of option pricing.
We specify the environment details as well as how the agent is parametrized
and trained. We provide a few numerical experiments which compare
the learned optimal hedging against the BS predictions. We also study
how the factors affect the optimal price and position. We hope that
this work can inspire and motivate more research into the cross-discipline
study between finance and machine learning.

We would like to discuss a few possible future directions, starting
from the financial aspects. We focus on European call options in this
report, but in general, it shall also apply to the put option or other
options that can only exercise at the terminal time. It is not clear
whether the QLBS or RLOP model can generalize to the American or even
swing options that are useful in electricity trading. Besides, the
transaction cost model proposed in Sec. \ref{subsec:Trading-Cost}
can be greatly improved to allow fine models on the order density,
or even influence on the mid price.

There are also a few open questions on the mathematics side. It could
be beneficial to analyze the theoretical optimal hedging strategy
in the discrete time model. We have provided an analogous to the stochastic
integral in the earlier sections, which might help in analyzing the
temporal discretization error. Besides, the modified QLBS reward function
(Eqn. \ref{eq:qlbs-modified-V}) is no longer time translational invariant,
so one has to train a new model if the maturity time of the option
changes. It would be a great improvement if one finds an equivalent
or a similar formulation where the value function is translational
invariant while giving homogeneous rewards over time.

We also have a few comments on the learning/sampling algorithms. One
common drawback of the two methods is that the computational complexity
scales quadratically as the maturity time $T$. The modified QLBS
model requires computing a conditional expectation at each time step,
which is not of constant cost. The RLOP model stacks $T$ pricing
tasks together, each lasting one more period than the previous one,
leading to a quadratic cumulative cost. Another issue is that we use
REINFORCE with baseline as the learning algorithm, which can be replaced
by the actor-critic (or even the soft version) to improve performance.
More time is appreciated since we have limited time in the project
to try different possibilities.

There are controversial conclusions as far as what the numerical experiments
show. In general, the agents learn to price a higher price if the
current price $S_{t}$ is higher, or the risk aversion parameter $\lambda$
(or the friction parameter $\epsilon$) is larger. However, the two
models predict the optimal hedging strategy differently, especially
when large friction is anticipated. The modified QLBS agent prefers
to hedge more than the frictionless case while the RLOP agent inclines
to hedge less. This discrepancy might be the key to understanding
the difference between these two models and eventually the option
pricing problem.

\bibliographystyle{siam}
\bibliography{ref}

\end{document}